\documentclass[
  journal=pasa,
  manuscript=research-paper,
  year=2023,
  volume=37,
]{cup-journal}

\usepackage{amsmath, amssymb}
\usepackage[nopatch]{microtype}
\usepackage{booktabs}

\usepackage{hyphenat}
\usepackage{hyperref}
\usepackage{ulem}
\usepackage{amssymb}
\usepackage{comment}

\hypersetup{colorlinks=true,linkcolor=blue,urlcolor=blue,citecolor=blue,anchorcolor=blue}

\title{The Third Konus-Wind Catalog of Short Gamma-Ray bursts}

\author{Alexandra L. Lysenko}
\affiliation{Ioffe Institute, Polytekhnicheskaya, 26, St. Petersburg, 194021 -  Russian Federation}
\email[A. Lysenko]{alexandra.lysenko@mail.ioffe.ru}

\author{Dmitry S. Svinkin}
\affiliation{Ioffe Institute, Polytekhnicheskaya, 26, St. Petersburg, 194021 -  Russian Federation}
\author{Dmitry D. Frederiks}
\affiliation{Ioffe Institute, Polytekhnicheskaya, 26, St. Petersburg, 194021 -  Russian Federation}

\author{Anna V. Ridnaia}
\affiliation{Ioffe Institute, Polytekhnicheskaya, 26, St. Petersburg, 194021 -  Russian Federation}

\author{Anastasia E. Tsvetkova}
\affiliation{Dipartimento di Fisica, Universit\`{a} degli Studi di Cagliari, SP Monserrato-Sestu, km 0.7, I-09042 Monserrato, Italy}
\alsoaffiliation{Ioffe Institute, Polytekhnicheskaya, 26, St. Petersburg, 194021 -  Russian Federation}

\author{Mikhail V. Ulanov}
\affiliation{Ioffe Institute, Polytekhnicheskaya, 26, St. Petersburg, 194021 -  Russian Federation}

\keywords{gamma-ray bursts (629), magnetars (992), catalogs (205)} 

\begin{document}

\begin{abstract}
In this catalog, we present the results of a systematic study of 199 short gamma-ray bursts (GRBs) detected by Konus-\textit{Wind} between 2011 January~1 and 2021 August~31. 
The catalog extends the Second Catalog of short gamma-ray bursts covering the period 1994--2010 by ten years of data. 
The resulting Konus-\textit{Wind} short GRB sample includes 494 bursts.
From temporal and spectral analyses of the sample, we provide the burst durations, spectral lags, estimates of the minimum variability time scales, rise and decay times, the results of spectral fits with three model functions, the total energy fluences, and the peak energy fluxes of the bursts. 
We present statistical distributions of these parameters for the complete set of 494 short gamma-ray bursts detected in 1994--2021.
 We discuss in detail the properties of the bursts with extended emission in the context of the whole short GRB population.
Finally, we consider the results in the context of the Type I (merger-origin)/Type II (collapsar-origin) classification, and discuss the extragalactic magnetar giant flare subsample.
\end{abstract}

\section{\label{sec_intro} Introduction}
Cosmological gamma-ray bursts (GRBs) are thought to be related to at least two distinct classes of catastrophic events: the merger of binary compact objects, such as two neutron stars or a neutron star and a black hole, and the core collapse of a massive star.  
The former may produce a short-duration GRB, with a duration less than $\sim 2$~s, the so called Type~I GRB; while the latter occasionally produce typically long-duration GRB, showing softer spectrum and non-negligible spectral lag, Type~II GRBs \citep[see, e.g.,][for more information on the Type~I/II classification scheme]{Zhang_2006Natur_444_1010, Zhang_2009ApJ_703_1696}. 

In fact, the duration distributions of Type~I and Type~II GRBs significantly overlap. 
The shortest Type~II burst discovered so far~--- GRB\,200826A~\citep{Ahumada2021} had a duration of $\lesssim 1$\,s, while GRB\,170817A, the counterpart of the gravitational-wave event GW\,170817 and kilonova AT2017gfo from a binary neutron star merger~\citep{Abbott_2017ApJ_848L_12}, had a duration of $\sim 2$~s. 
Moreover, for a number of nearby short/hard GRBs it was possible to obtain deep upper limits on both supernova and kilonova emission, \citep[see, e.g.,][]{Ferro2023}, which illustrates the complexity of physical  classification of short GRBs.
On the other hand, a number of long-duration GRBs showed afterglow features similar to kilonova emission produced by a binary merger, in particular two recent bursts: GRB\,211211A~\citep{Rastinejad_2022Natur_612_223, Troja_2022Natur_612_228, Barnes_2023ApJ_947_55} and GRB\,230307A~\citep{Levan_2023NatAs_133,Levan_2023arXiv230702098, Dichiara_2023arXiv230702996}.
These bursts show a short initial episode followed by a bright, tens of seconds-long main phase, which may have a similar origin to the so-called extended emission (EE) observed in a fraction of short GRBs (sGRBs). 
The EE is a weaker emission component that follows the short initial pulse (IP) has been observed in a fraction of sGRBs by various experiments: 
\textit{CGRO}-BATSE~\citep{Burenin_2000AstL, Norris_and_Bonnel_2006ApJ, Bostanci_2013MNRAS}, 
Konus-\textit{Wind} ~\citep[KW;][]{Mazets2002, Svinkin2016}, 
\textit{INTEGRAL}-SPI-ACS~\citep{Minaev_2010AstL}, 
\textit{Swift}-BAT~\citep{Norris_2011ApJ_735_23, Sakamoto_2011ApJS}, \textit{Fermi}-GBM~\citep{Kaneko_2015MNRAS}, and \textit{AGILE}-MCAL~\citep{Ursi2022}. 

In addition to cosmological GRBs, the observed sGRB population includes magnetar giant flares (MGFs) in nearby galaxies~\citep{Svinkin2021,Burns2021}, which may be identified by localization and distinct temporal and spectral evolution.

Thus, a detailed analysis of a large sample of sGRBs, including candidates for sGRBs with EE, may contribute significantly to the understanding of both the collapsar and the merger origin scenarios.

In this catalog, we update the results presented in the Second sGRB Catalog \citep[][hereafter S16]{Svinkin2016} to include the analysis of 199 sGRBs detected by KW between 2011 January~1 to 2021 August~31 (494 sGRBs in total). The burst sample selection criteria and localizations are presented in \cite{Svinkin2019,Svinkin2022}.

This catalog is organized as follows. We start with a short description of the KW instrument details in Section~\ref{sec_kw}. In Section~\ref{sec_sample}, we provide details of the KW sGRB sample. We describe the light curve and spectral analysis procedures and present the results in Section~\ref{sec_analysis}. In Section~\ref{sec_discussion}, we discuss the results. In Section~\ref{sec_conclusions}, we give a summary and our conclusions.

Throughout the paper, all errors are reported at the 68\% confidence level (CL).

\section{\label{sec_kw} Konus-\textit{Wind} }

The KW~\citep{Aptekar1995} spectrometer was launched on board the NASA \textit{Wind} spacecraft in 1994~November and has operated until the present day. 
The instrument consists of two identical NaI(Tl) detectors S1 and S2, each with a 2$\pi$ field of view and a nominal energy band of 10 keV - 10 MeV. 
The detectors are mounted on opposite faces of the rotationally stabilized spacecraft,
such that one detector (S1) points toward the south ecliptic pole, thereby observing the southern ecliptic hemisphere, while the other (S2) observes the northern ecliptic hemisphere.

KW operates in two modes: waiting and triggered. 
In the waiting mode, light curves in three wide energy bands with nominal boundaries  G1~(13--50\,keV), G2~(50--200\,keV), and G3~(200--760\,keV) are recorded with time resolution of 2.944\,s. 
The switching to the triggered mode occurs at a statistically significant count rate increase in the G2 band. 
In the triggered mode, light curves are recorded in the same bands G1, G2, and G3 from $-0.512$\,s to 229.632\,s relative to the trigger time with a varying time resolution.
The time resolution is 2\,ms for the interval between $-0.512$\,s and 0.512\,s, 16\,ms for the 0.512--33.280\,s, 64\,ms for the 33.280--98.816\,s relative to the trigger time, and 256\,ms for the remaining triggered time history. 
In the triggered mode, 64 multichannel spectra are measured starting from the trigger time by two pulse-height analyzers PHA1 (63 channels, nominal boundaries 13--760\,keV) and PHA2 (60 channels, nominal boundaries 250\,keV--10\,MeV). 
The first four spectra are measured with a fixed accumulation time of 64\,ms
in order to study short bursts. 
For the subsequent 52 spectra, an adaptive system determines the accumulation times, which may vary from 0.256\,s to 8.192\,s depending on the current count rate in the G2 energy band (the higher count rates produces the shorter accumulation times). 
The last 8 spectra have accumulation time of 8.192\,s each. 
As a result, the duration of spectral measurements varies from 79.104\,s, to 491.776\,s.

The detector response matrix (DRM) was calculated using GEANT4 toolkit \citep{Agostinelli2003, Allison2006, Allison2016}. 
In this work, as compared to S16, we used an updated DRM that characterizes the detector response more accurately. 
We discuss the impact of the DRM update on results further.

The detector gain slowly decreases with time. 
As of 2021, the detector energy ranges have shifted from the nominal 13\,keV--10\,MeV to 28\,keV--20\,MeV (S1) and 22\,keV--16\,MeV (S2), and the light curve and the spectral measurement bands have shifted accordingly.
A more detailed instrument description can be found in S16 and~\cite{Lysenko2022}.

\section{\label{sec_sample} The sGRB Sample}

\begin{figure}
    \centering
    \includegraphics[width=0.99\textwidth]{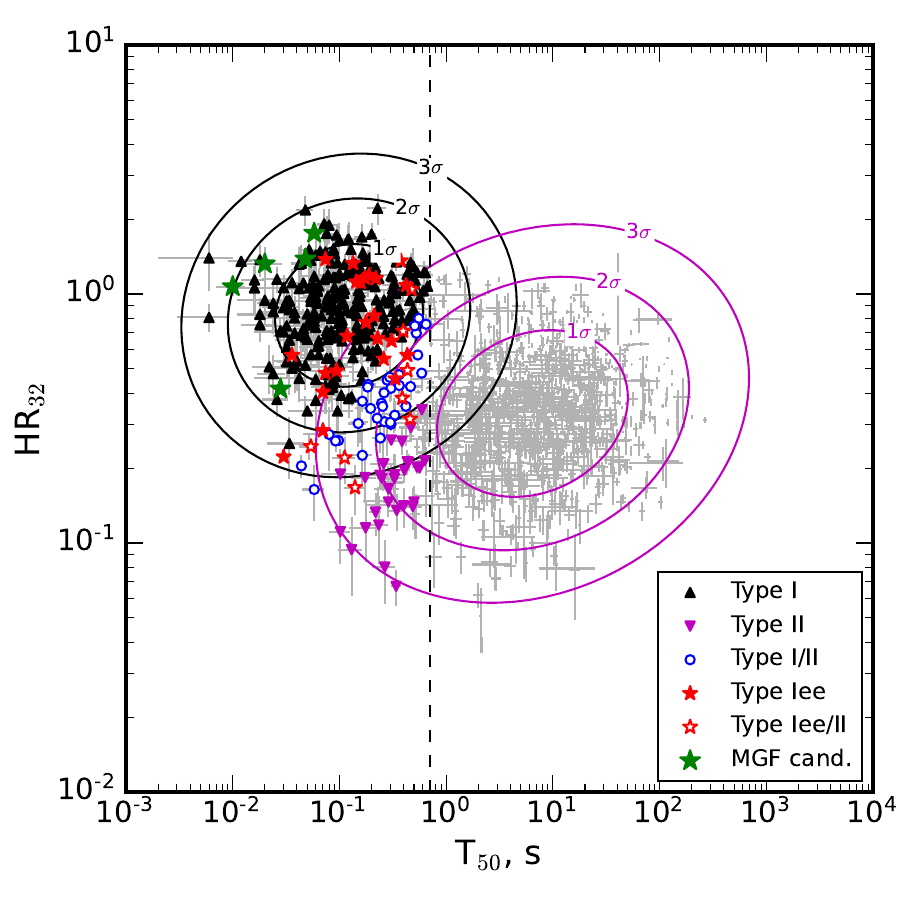}
    \caption{Hardness-duration distribution for 3398 KW GRBs. The distribution is fitted with a sum of two Gaussian distributions. The contours correspond to $1\sigma$, $2\sigma$, and $3\sigma$ cumulative probability for short-hard (black) and long-soft (magenta) distributions. The vertical dashed line denotes the boundary at $T_{50}$ = 0.7\,s between short and long GRBs. The types for sGRBs are shown in colors: Type~I (black triangles), Type~I/II (blue circles), Type~II (magenta inverted triangles), Type~Iee (filled red stars), Type~Iee/II (empty red stars), MGF candidates (green stars). The remaining GRBs are plotted as gray crosses.}
    \label{fig_t50_vs_hr32}
\end{figure}

\begin{table*}
\begin{threeparttable}
\caption{ KW sGRB Observation Details (Sample~II)}.
\label{tab_loc}
\begin{tabular}{ccccrrccc}
\toprule
\headrow Designation & KW & Name\tnote{a} & Detector & R.A. & Dec. & Incident Angle & Comment\tnote{b} & Type \\
 & Trigger Time (UT) & & & ($^\circ$) & ($^\circ$) & ($^\circ)$ &  &  \\
\midrule
GRB20110212\_T47551 & 13:12:31.101 &  & S1 & 348.236 & -72.931 & 31$_{-1}^{+4}$ & 2 & I\\
GRB20110221\_T18490 & 05:08:10.017 &  & S2 & 316.305 &  -5.895 & 80$_{-1}^{+1}$ & 2 & I\\
GRB20110323\_T57460 & 15:57:40.228 &  & S1 &  10.195 &  -4.350 & 82$_{-6}^{+5}$ & 2 & I\\
GRB20110401\_T79461 & 22:04:21.937 & GRB110401A & S2 & 264.968 &  24.902 & 42$_{-3}^{+3}$ & 2 & I/II\\
GRB20110510\_T80844 & 22:27:24.326 &  & S1 & 334.020 & -43.873 & 59$_{-2}^{+2}$ & 2 & I/II\\
\bottomrule
\end{tabular}
\begin{tablenotes}[hang]
\item[]Table note
\item[a]As provided in the GCN circulars, if available.
\item[b]1~--- the burst was detected by an imaging instruments (the incident angle error is negligible and not given); 2~--- the burst was localized by IPN; R.A. and Dec. correspond to the most probable source location; 3~--- the burst was localized by IPN to a long segment, R.A. and Dec. are not given, the source ecliptic latitude estimate is used for the incident angle calculation; 4~--- the burst was localized to a large region, R.A. and Dec. are not given, the incident angle is set at 60$^{\circ}$.

(This table is available in its entirety in machine-readable form via \url{http://www.ioffe.ru/LEA/shortGRBs/Catalog3/}.)
\end{tablenotes}
\end{threeparttable}
\end{table*}

Between the start of the operation on 1994 November~12 and 2021 August~31, KW triggered on 3397 GRBs.
During the interval covered with this catalog, KW detected 1365 GRBs. 
From the updated analysis~\citep{Svinkin2019} of KW burst durations $T_{50}$ and $T_{90}$\footnote{$T_{50}$ and $T_{90}$ durations are the time intervals which contain from 25\% to 75\% and from 5\% to 95\% of the total burst count fluence, respectively \citep{Kouveliotou_1993ApJ}}, we adopted $T_{50} = 0.7$\,s as the boundary between short and long KW GRBs, which yielded 198\footnote{As compared to \cite{Svinkin2022} one burst, namely, GRB20150702\_T86198, initially classified as sGRB with EE, was excluded from the sGRB sample, since its initial pulse has $T_{50}>0.7$\,s.} sGRBs detected between 2011 January and 2021 August. 

Previous studies suggest that some of the sGRBs can, in fact, be initial pulses of magnetar giant flares (MGFs) from nearby galaxies \citep[see, e.g.,][]{Svinkin2015, Burns2021}. 
The sample of KW sGRBs detected up to August 2021 contains only four bursts whose localization regions overlap nearby galaxies and which can thus be interpreted as extragalactic MGF candidates.
These bursts are GRB\,051103 in the M81/M82 group of galaxies \citep{Frederiks2007}, GRB\,070201 in the Andromeda galaxy M31~\citep{Mazets2008}, GRB\,070222 in the M83 galaxy~\citep{Burns2021}, and GRB\,200415A in the Sculptor galaxy (NGC~253)~\citep{Svinkin2021}. 
The first two bursts were discussed in S16. 
To update the sample of KW-detected MGF candidates (hereafter, we will refer to them simply as MGFs), we consider also the recent GRB\,231115A, detected on 2023, November, 15 \citep{Mereghetti2024, Minaev2024, Frederiks2023b, Trigg2024a}, a candidate MGF associated with M82 galaxy.

Hereafter we denote the KW sGRB sample detected in 1994--2010 as Sample~I, the KW sGRB sample detected in 2011--2021 plus the recent MGF GRB\,231115A as Sample~II, and Sample~I plus Sample~II as the Full sGRB Sample.

We searched for sGRBs with EE candidates in Sample~II using the criteria similar to those used in S16: short ($T_{50} \leq 0.7$\,s) IP followed by a weaker emission without separate intense peaks and without prominent spectral evolution. 

The bursts were classified into physical types using a two 2D Gaussian component fit to the hardness-duration distribution, in a way similar to S16, \citet[][hereafter T17]{Tsvetkova2017}, \cite{Svinkin2019}, see Figure~\ref{fig_t50_vs_hr32}. 
Hereafter all Figures refer to the Full sGRB Sample.
The burst spectral hardness, $HR_\mathrm{32}$ was calculated using the ratio of counts in the
G3 and G2 bands accumulated during the total burst duration ($T_\mathrm{100}$, see Section~\ref{sec_temp_analysis}). 
The calculation of $HR_\mathrm{32}$ accounts for the KW gain drift (see S16).
According to this classification Sample~II includes 147 Type~I GRBs (merger origin; 76\%), 22 Type~II GRBs (collapsar origin; 11\%), for 17 sGRBs the type is uncertain (I/II; 8\%).
The classification of sGRBs that show extended emission was based on the IP parameters: six Iee bursts (Type I with EE), and one Iee/II (the type is uncertain: Iee or II).

Burst localization is essential for GRB spectral analysis, but KW has only coarse autonomous localization capability.
In cases where the position of a GRB is not available from an imaging instrument (e.g., \textit{Swift}-BAT), the source localization can be derived using the InterPlanetary network (IPN) triangulation (see, e.g.,~\citealp{Hurley_2013EAS}). 
The localizations of the KW sGRBs are reported in~\cite{Palshin_2013ApJS} and \cite{Svinkin2022}.

Table~\ref{tab_loc} lists the 199 sGRBs of Sample~II. The first column gives
the burst designation in the form {\scriptsize \mbox{``GRBYYYYMMDD\_Tsssss''}}, where YYYYMMDD is the
burst date, and sssss is the KW trigger time $T_0$ (UT) truncated to integer seconds (note that due to Wind's large distance from Earth, this trigger time can differ by up to $\sim 7$\,s from the near-Earth instrument detection times; see \citealt{Svinkin2022}). The second column gives the KW trigger time in the standard time format. 
The ``Name'' column specifies the GRB name as provided in the Gamma-ray Burst Coordinates Network circulars\footnote{https://gcn.nasa.gov/circulars} if available. 
The ``Detector'' column specifies the triggered detector. 
The columns ``R.A.'' and ``Dec.'' give sGRB localization (see Footnote (b) to Table~\ref{tab_loc}).
The next column provides the angle between the GRB direction and the detector axis (the incident angle). 
The last but one column contains localization-specific notes and the last column specifies the burst type.

\section{\label{sec_analysis} Data Analysis and Results}

\subsection{\label{sec_temp_analysis} Durations and spectral lags}

\begin{table*}
\begin{threeparttable}
\caption{Durations, Spectral Lags, and Classification (Sample~II)}.
\label{tab_temporal}
\begin{tabular}{ccccccccc}
\toprule
\headrow Designation & $t_{0}$\tnote{a} & $T_{100}$ &$T_{50}$ & $T_{90}$ & Type & $\tau_\mathrm{lag21}$ & $\tau_\mathrm{lag32}$ & $\tau_\mathrm{lag31}$ \\
\headrow  & (s) &(s) & (s) & (s) &  & (ms) & (ms) & (ms) \\
\midrule
GRB20110212\_T47551 & $-0.018$ & 0.088 & 0.018$\pm$0.006 &  0.052$\pm$0.027 & I    &    6$\pm$4  & 2$\pm$2 & 9$\pm$12 \\
GRB20110221\_T18490 & $-0.382$ & 0.426 & 0.056$\pm$0.009 &  0.388$\pm$0.015 & I    &  & 2$\pm$4  &  \\
GRB20110323\_T57460 & $-0.054$ & 0.066 & 0.042$\pm$0.010 &  0.064$\pm$0.007 & I    &  & 4$\pm$13 &  \\
GRB20110401\_T79461 & $-0.214$ & 0.790 & 0.508$\pm$0.037 &  0.694$\pm$0.039 & I/II & $-9\pm32$ & 4$\pm$10 &  \\
GRB20110510\_T80844 & $-0.470$ & 0.950 & 0.364$\pm$0.079 &  0.840$\pm$0.074 & I/II &   &  &   \\
\bottomrule
\end{tabular}
\begin{tablenotes}[hang]
\item[]Table note
\item[a]Relative to the trigger time. \\
(This table is available in its entirety in machine-readable form via \url{http://www.ioffe.ru/LEA/shortGRBs/Catalog3/}.)
\end{tablenotes}
\end{threeparttable}
\end{table*}

\begin{figure*}
    \centering
    \includegraphics[width=0.8\textwidth]{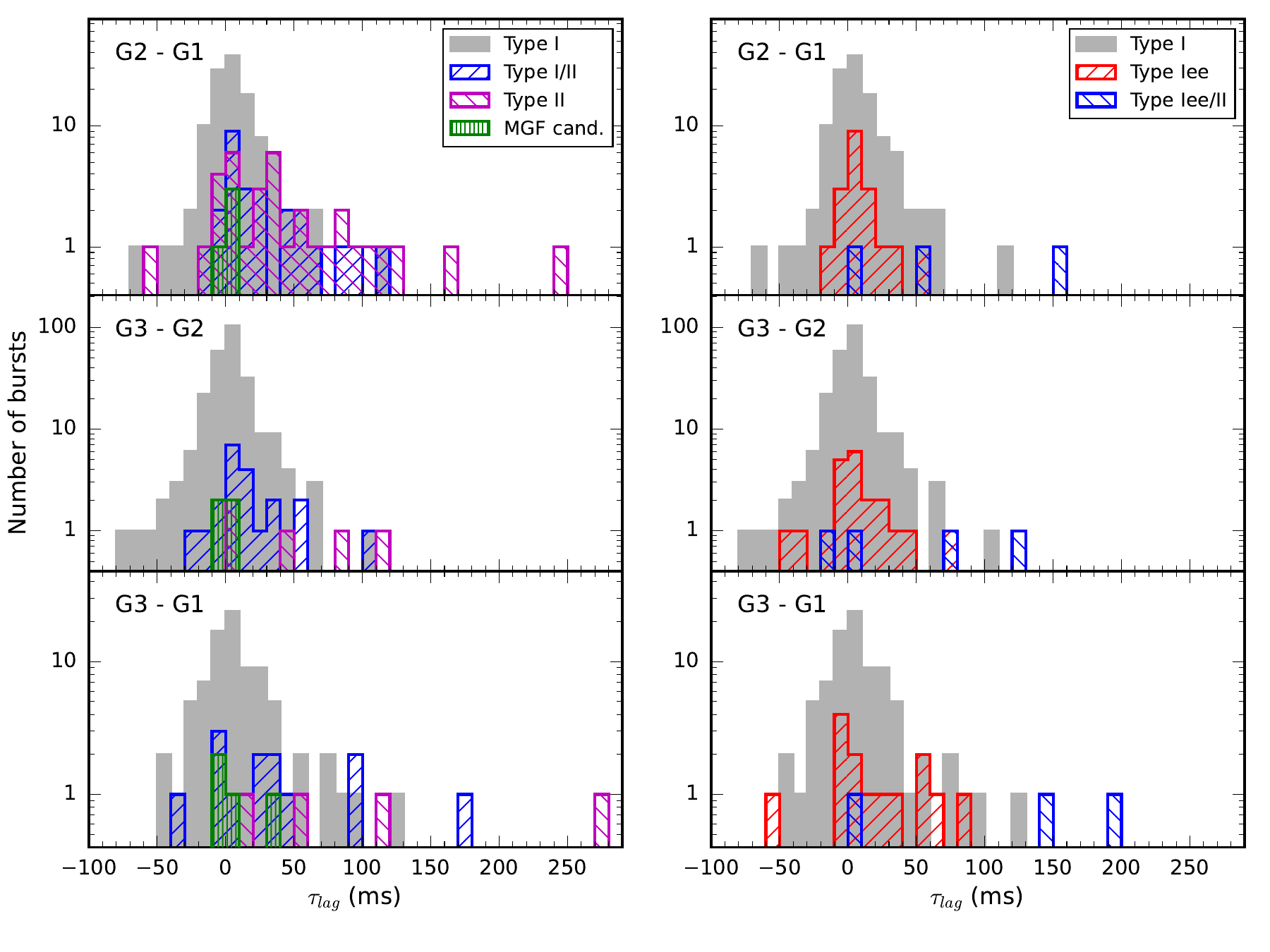}
    \caption{Spectral lag distributions. Left~--- distributions for Type~I bursts (gray), Type~I/II bursts (blue), Type~II bursts (magenta), and MGF candidates (green). Right~--- distributions for Type~I bursts (gray), Type~Iee bursts (red), and Type~Iee/II bursts (blue). In each column, the panels correspond to the following pairs of energy bands: G2 and G1 (top), G3 and G2 (middle), and G3 and G1 (bottom).}
    \label{fig_lags}
\end{figure*}

The burst durations $T_{100}$ (the total duration estimated using the $5\sigma$ threshold), $T_{90}$, and $T_{50}$ are calculated following the procedure described in~\citet{Svinkin2019}.

The spectral lag is a quantitative measure of spectral evolution of GRBs. 
A positive lag corresponds to the delay of emission in a softer energy band relative to a harder one.
It was shown that sGRBs both with and without EE have negligible spectral lag~\citep{Norris_2001conf,Norris_and_Bonnel_2006ApJ}.
Thus, the spectral lag can be used as an additional classification parameter.

We report spectral lags between lightcurves in the G2 and G1 ($\tau_\mathrm{lag21}$), G3 and G2 ($\tau_\mathrm{lag32}$), and G3 and G1 ($\tau_\mathrm{lag31}$) energy bands. 
We calculated the lags using the cross-correlation function (CCF) similar to that used by \cite{Link1993}, \cite{Fenimore1995}, and \cite{Norris2000}. 
We defined lag as the position of the maximum of the second order polynomial fit to CCF near its peak.  
To estimate lag uncertainties, we used Monte Carlo simulations of the burst light curves in each energy band. For each simulation, we added Poisson noise to both light curves according to the count rates and calculated a spectral lag.
The resulting lag was defined as the median of the simulated lag distribution and the lag uncertainty was defined as the half-width of the 68\% CL.
For each burst, the time interval for cross-correlation, the temporal resolution of the light curve, and the CCF fitting interval were individually adjusted to account for the duration and the intensity of the event.
For the bursts with poor count statistics in one or two channels, the corresponding lags were not calculated.
For bursts with EE, spectral lags were calculated for the IP only. 

Table~\ref{tab_temporal} contains the burst durations, the spectral lags, and the
classification for Sample~II. The first column gives the burst designation. 
The following four columns contain the start of the $T_{100}$ interval $t_0$ (relative to $T_0$), $T_{100}$, $T_{90}$, and $T_{50}$. 
The next column gives the Type~I--II classification and the last three columns contain spectral lags $\tau_\mathrm{lag21}$, $\tau_\mathrm{lag32}$, and $\tau_\mathrm{lag31}$.

Figure~\ref{fig_lags} presents spectral lag distributions. 
For Type~I bursts the measured lags are distributed around zero, which is in agreement with the previous study (S16).
We note that longer lags for Type~I bursts tend to have large uncertainties (for further discussion see Section~\ref{sec_discussion}).

\subsection{Minimum variability time scale, rise and decay times}

\begin{table}
\begin{threeparttable}
\caption{ Minimum variability time scale, rise and decay times for the Full sGRB Sample.}
\label{tab_bb}
\begin{tabular}{ccccc}
\toprule
\headrow Designation & $\delta T$ & $\tau_\mathrm{rise}$ & $\tau_\mathrm{decay}$ & $\tau_\mathrm{rise}$/$\tau_\mathrm{decay}$ \\
\headrow  & (ms) & (ms) & (ms) & \\
\midrule
GRB19950211\_T08697 &  8 &  65 &  121 & 0.537 \\
GRB19950210\_T08424 & 78 &  39 &  127 & 0.307 \\
GRB19950414\_T40882 &  74 & 131 &   37 & 3.540 \\
GRB19950503\_T66971 &  52 & 245 &  105 & 2.330 \\
GRB19950520\_T83271 &  76 & 166 &  214 & 0.776 \\
\bottomrule
\end{tabular}
(This table is available in its entirety in machine-readable form via \url{http://www.ioffe.ru/LEA/shortGRBs/Catalog3/}.)
\end{threeparttable}
\end{table}

To estimate the minimum variability time scale, rise, and decay times we used the Bayesian block decomposition \citep{Scargle2013} of the burst light curves. 
This decomposition segments a light curve into intervals, blocks, and within each block the source count rate is assumed to be constant at a given significance level, with variations caused by random fluctuations\footnote{We used our own implementation for Bayesian block decomposition. The source code is available via \url{https://github.com/dsvinkin/b_blocks}}. 
We performed the segmentation on the sum of the observed count rates in the G2 and G3 channels at the significance level corresponding to $\sim 5\sigma$.
The minimum variability time scale $\delta T$ was evaluated for each burst as the duration of the shortest block. 
The peak time was defined as the center time of the block with the maximum count rate. 
The rise time $\tau_\mathrm{rise}$ was estimated as the difference between the peak time and the beginning time of the first block of the burst, and the decay time $\tau_\mathrm{decay}$~--- as the difference between the end time of the last block and the peak time. 
This approach, however, does not account for the multi-peaked structure observed for a number of sGRBs.
For bursts, described with a single block, rise and decay times were not estimated. 

\begin{figure*}
    \centering
    \includegraphics[width=0.9\textwidth]{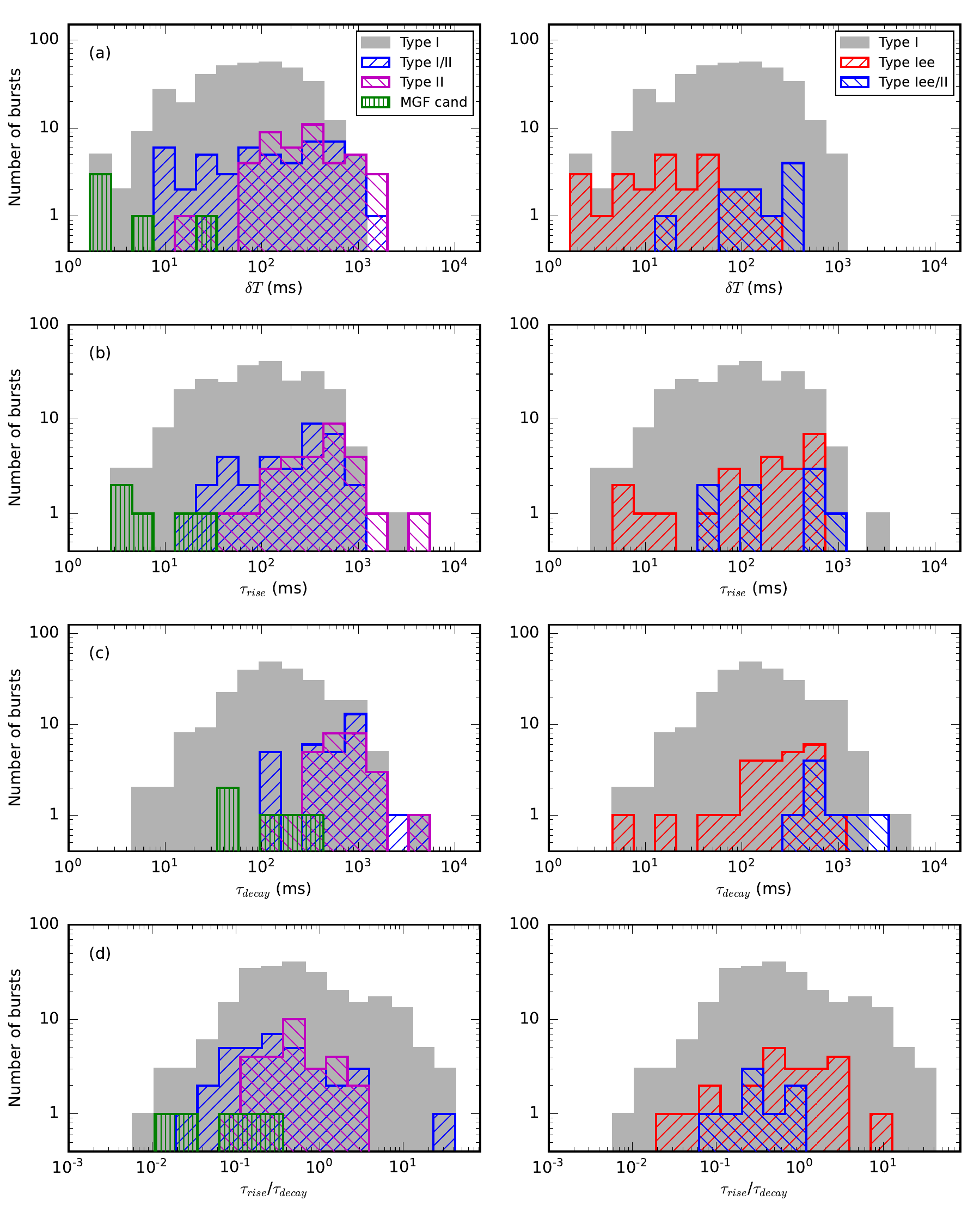}
    \caption{Burst variability time scales derived for the G2+G3 channel light curves. The left column presents distributions for Type~I bursts (gray), Type~I/II bursts (blue), Type~II bursts (magenta), and MGF candidates (green). The right column~--- distributions for Type~I bursts (gray), Type~Iee bursts (red), and Type~Iee/II bursts (blue). Each row shows the distributions for the following parameters: (a) Minimum variability time scale $\delta T$; (b) Rise time $\tau_\mathrm{rise}$; (c) Decay time $\tau_\mathrm{decay}$; (d) Rise-to-decay time ratio.}
    \label{fig_bb}
\end{figure*}

Table~\ref{tab_bb} lists $\delta T$, $\tau_\mathrm{rise}$, $\tau_\mathrm{decay}$ and the ratio $\tau_\mathrm{rise}$/$\tau_\mathrm{decay}$ for the Full sGRB sample. 
Figure~\ref{fig_bb} presents distributions of these values for different sGRBs types.
The distribution of $\delta T$ peaks at $\sim 100$\,ms and also has a minor separated peak at 2\,ms.
Given that 2\,ms is the finest time resolution available for KW, this small peak corresponds to the bursts with $\delta T \le 2$\,ms.
The shortest $\delta T$ are typical for MGFs: three out of five MGFs have $\delta T \le  2$\,ms. 
MGFs are also characterized by fast rise times: for three out of five bursts $\tau_\mathrm{rise}< 8$\,ms. 
Fast time variability is typical for Type~I and Type~Iee bursts, while Type~II bursts demonstrate significantly longer scales (for further discussion see Section~\ref{sec_discussion}).

\subsection{\label{ssec_spec_analysis} Spectral Analysis}

\begin{table*}
\begin{threeparttable}
\caption{Spectral Parameters (Multichannel Spectra) for Sample~II}.
\label{tab_mult}
\begin{tabular}{cccccccccc}
\toprule
\headrow Designation & Spec. & $T_\mathrm{start}$\tnote{a} & $\Delta T$ & Model\tnote{b} & $\alpha$ & $\beta$ & $E_{p}$ & Flux\tnote{c} & $\chi^2$/d.o.f. \\
\headrow  & Type & (s) & (s) &  &  &  & (keV) &  & (Prob.) \\
\midrule
GRB20110221\_T18490 & i,p & 0.000 & 0.064 & *PL & -1.2$_{-0.09}^{+0.09}$ &  &  & 36.92$_{-7.66}^{+8.20}$ & 22.5/17 (1.7e-01) \\
 &  &  &  & CPL & -0.67$_{-0.29}^{+0.35}$ &  & 1926$_{-600}^{+1572}$ & 20.85$_{-5.08}^{+7.35}$ & 18.7/16 (2.8e-01) \\
 &  &  &  & BAND & -0.68$_{-0.25}^{+0.32}$ & -1.61\tnote{d} & 1932$_{-742}^{+1562}$ & 20.91$_{-2.87}^{+5.27}$ & 18.7/15 (2.3e-01) \\
GRB20110526\_T61739 & i,p & 0.000 & 0.256 & PL & -1.52$_{-0.06}^{+0.05}$ &  &  & 7.33$_{-1.20}^{+1.32}$ & 55.1/21 (6.8e-05) \\
 &  &  &  & *CPL & -0.58$_{-0.24}^{+0.32}$ &  & 474$_{-115}^{+161}$ & 3.24$_{-0.60}^{+0.71}$ & 22.4/20 (3.2e-01) \\
 &  &  &  & BAND & 0.09\tnote{e} & -1.85\tnote{d} & 134$_{-38}^{+177}$ & 4.5$_{-1.08}^{+1.22}$ & 22.3/19 (2.7e-01) \\
GRB20110529\_T02920 & i,p & 0.000 & 0.256 & PL & -1.37$_{-0.03}^{+0.03}$ &  &  & 18.48$_{-1.72}^{+1.73}$ & 89.6/48 (2.5e-04) \\
 &  &  &  & CPL & -0.85$_{-0.11}^{+0.14}$ &  & 1176$_{-317}^{+445}$ & 10.84$_{-2.03}^{+2.44}$ & 35.9/47 (8.8e-01) \\
 &  &  &  & *BAND & -0.37$_{-0.24}^{+0.27}$ & -1.88$_{-0.18}^{+0.12}$ & 451$_{-102}^{+160}$ & 13.45$_{-2.03}^{+2.08}$ & 30.7/46 (9.6e-01) \\
GRB20110705\_T13029 & i,p & 0.000 & 0.192 & *CPL & -0.19$_{-0.12}^{+0.14}$ &  & 1005$_{-107}^{+120}$ & 26.1$_{-2.36}^{+2.49}$ & 23.5/30 (7.9e-01) \\
 &  &  &  & BAND & -0.17$_{-0.14}^{+0.19}$ & -2.80\tnote{d} & 981$_{-159}^{+121}$ & 26.63$_{-2.72}^{+3.11}$ & 23.5/29 (7.6e-01) \\
GRB20110802\_T55156 & i,p & 0.000 & 0.256 & PL & -1.16$_{-0.05}^{+0.05}$ &  &  & 21.15$_{-2.47}^{+2.49}$ & 42.8/26 (2.0e-02) \\
 &  &  &  & *CPL & -0.64$_{-0.16}^{+0.18}$ &  & 3795$_{-842}^{+1306}$ & 21.55$_{-3.36}^{+3.35}$ & 25.0/25 (4.6e-01) \\
 &  &  &  & BAND & -0.64$_{-0.16}^{+0.19}$ & -1.90\tnote{d} & 3676$_{-918}^{+1270}$ & 21.65$_{-3.27}^{+3.30}$ & 24.8/24 (4.2e-01) \\
\bottomrule
\end{tabular}
\begin{tablenotes}[hang]
\item[]Table note
\item[a]Relative to the trigger time.
\item[b]The best-fit model is indicated by the asterisk.
\item[c]In units of 10$^{-6}$ erg cm$^{-2}$ s$^{-1}$.
\item[d]Upper limits.
\item[e]Lower limits. \\
(This table is available in its entirety in machine-readable form via \url{http://www.ioffe.ru/LEA/shortGRBs/Catalog3/}.)
\end{tablenotes}
\end{threeparttable}
\end{table*}

For a typical KW sGRB, the time-integrated (TI) spectrum may be well characterized using a subset of the first four 64\,ms spectra measured from $T_0$ up to $T_0+0.256$\,s. 
For some bright, relatively-long events, we include also the 5th spectrum with an accumulation time up to 8.192\,s. 
The background spectrum for a burst without EE was usually taken starting from $\sim T_0+25$\,s with an accumulation time of about 100\,s. 
For about 52\% of the bursts in Sample~II, a major fraction of the total count fluence was accumulated before the trigger time, and is not covered by the multichannel spectra. 
For these bursts, we use a three-channel spectrum constructed from the time-integrated light curve counts in G1, G2, and G3 (see S16).

We analyzed a total of 94 multichannel and 105 three-channel TI spectra for Sample II.
Due to low counting statistics of the majority of KW sGRBs, we typically use the burst TI spectrum to calculate its total energy fluence ($S$) and peak energy flux ($F_\mathrm{peak}$). 
Only for 11 fairly intense GRBs, it was possible to derive $F_\mathrm{peak}$ from a dedicated, ``peak'' spectrum covering a narrow time interval near the peak count rate.

We chose three spectral models to fit the spectra of GRBs from our sample. 
These models were a simple power law (PL), an exponential cutoff power law (CPL), and the Band's GRB function (BAND; \citealt{Band1993}). All models are formulated in units of photon flux $f$ (photon~s$^{-1}$~cm$^{-2}$~keV$^{-1}$). The details of each model are presented below.

The power-law model
\begin{equation}
\label{eq_pl}
f_{\rm PL} \propto E^{\alpha}\,.
\end{equation}

The power-law with exponential cutoff 
\begin{equation}
\label{eq_cpl}
f_{\rm CPL} \propto E^{\alpha} {\rm exp}\left(-\frac{E(2+\alpha)}{E_{\rm p}}\right)\,.
\end{equation}

The Band's function:
\begin{equation}
\label{eq_band}
f_{\rm BAND} \propto 
\begin{cases}
E^{\alpha} {\rm exp}\left(-\frac{E(2+\alpha)}{E_{\rm p}}\right), E<\frac{E_{\rm p}(\alpha-\beta)}{2+\alpha} \\
E^{\beta} {\rm exp}\left (\beta-\alpha \right) \left[ \frac{E_{\rm p}(\alpha-\beta)}{2+\alpha}\right]^{\alpha-\beta}, E \geq \frac{E_{\rm p}(\alpha-\beta)}{2+\alpha}\,.
\end{cases}
\end{equation}

Here $E_{\rm p}$ is the peak energy of the $EF_E$ spectrum, $\alpha$ is the low-energy photon index, and $\beta$ is the photon index at higher energies.

Spectral fitting was performed in XSPEC~12.11.1 package~\citep{Arnaud1996} using the $\chi^2$ statistics.
For multi-channel spectra, to ensure the validity of the $\chi^2$ statistics, we grouped the energy channels to have at least 20 counts per channel. 
We used a model energy flux in the 10\,keV--10\,MeV band as the model normalization during a fit. 
The flux was calculated using the \texttt{cflux} convolution model in XSPEC.
The parameter errors were estimated using the XSPEC command \texttt{error} based on the change in the fit statistics ($\Delta \chi^2 = 1.0$), which corresponds to 68\%~CL.

The preferred model (the best-fit model) between PL, CPL and BAND was selected based on the $F$-test \citep{Bevington1969}.
As in S16, a model with an additional parameter was selected as the best-fit model when the decrease in $\chi^2$ corresponds to the probability of chance decrease $\leq 0.025$. 

The three-channel spectra were analyzed in a way similar to S16. 
In order not to overestimate the burst energetics, we limit the analysis to the CPL function.\footnote{BAND fits with four parameters are not available for a 3-channel spectrum.} Since the CPL fit to a three-channel spectrum has zero degrees of freedom, the parameter errors were estimated using the XSPEC \texttt{steppar} command.
For the cases where $\alpha$ or $E_{\rm p}$ were poorly constrained, typically, if
$E_{\rm p}$ is within G1 energy band or $E_{\rm p}$ is above the three-channel analysis range (i.e., $\gtrsim 1$~MeV), we report the parameter lower (or upper) limits.

\begin{table*}
\begin{threeparttable}
\caption{Spectral Parameters (Three-channel Spectra) for Sample~II.}
\label{tab_3chan}
\begin{tabular}{ccccccc}
\toprule
\headrow Designation & $T_{\rm start}$\tnote{a} & $\Delta T$\tnote{b} & Model & $\alpha$ & $E_{\rm p}$ & Flux\tnote{c}  \\
\headrow  & (s) & (s) &  &  & (keV) &  \\
\midrule
GRB20110212\_T47551 & -0.018 & 0.088 & CPL & -0.68$_{-0.20}^{+0.22}$ & 632$_{-144}^{+278}$ & 10.24$_{-1.72}^{+2.59}$ \\ 
GRB20110323\_T57460 & -0.054 & 0.066 & CPL &  0.23$_{-0.54}^{+0.80}$ & 416$_{-69}^{+107}$ & 8.62$_{-1.37}^{+1.55}$ \\
GRB20110401\_T79461 & -0.214 & 0.790 & CPL & -0.89$_{-0.21}^{+0.25}$ & 1131$_{-439}^{+2890}$ & 3.18$_{-0.88}^{+3.71}$ \\
GRB20110510\_T80844 & -0.470 & 0.950 & CPL & -0.42$_{-0.28}^{+0.35}$ & 266$_{-53}^{+63}$ & 1.27$_{-0.19}^{+0.20}$ \\
GRB20110514\_T79742 & -0.482 & 1.698 & CPL &  0.26$_{-0.36}^{+0.52}$ & 492$_{-58}^{+80}$ & 2.81$_{-0.23}^{+0.29}$ \\
\bottomrule
\end{tabular}
\begin{tablenotes}[hang]
\item[]Table note
\item[a]The burst start time relative to $T_{\rm 0}$.
\item[b]The burst total duration $T_{\rm 100}$.
\item[c]In units of 10$^{-6}$ erg cm$^{-2}$ s$^{-1}$. \\
(This table is available in its entirety in machine-readable form via \url{http://www.ioffe.ru/LEA/shortGRBs/Catalog3/}.)
\end{tablenotes}
\end{threeparttable}
\end{table*}

\begin{figure*}
    \centering
    \includegraphics[width=0.9\textwidth]{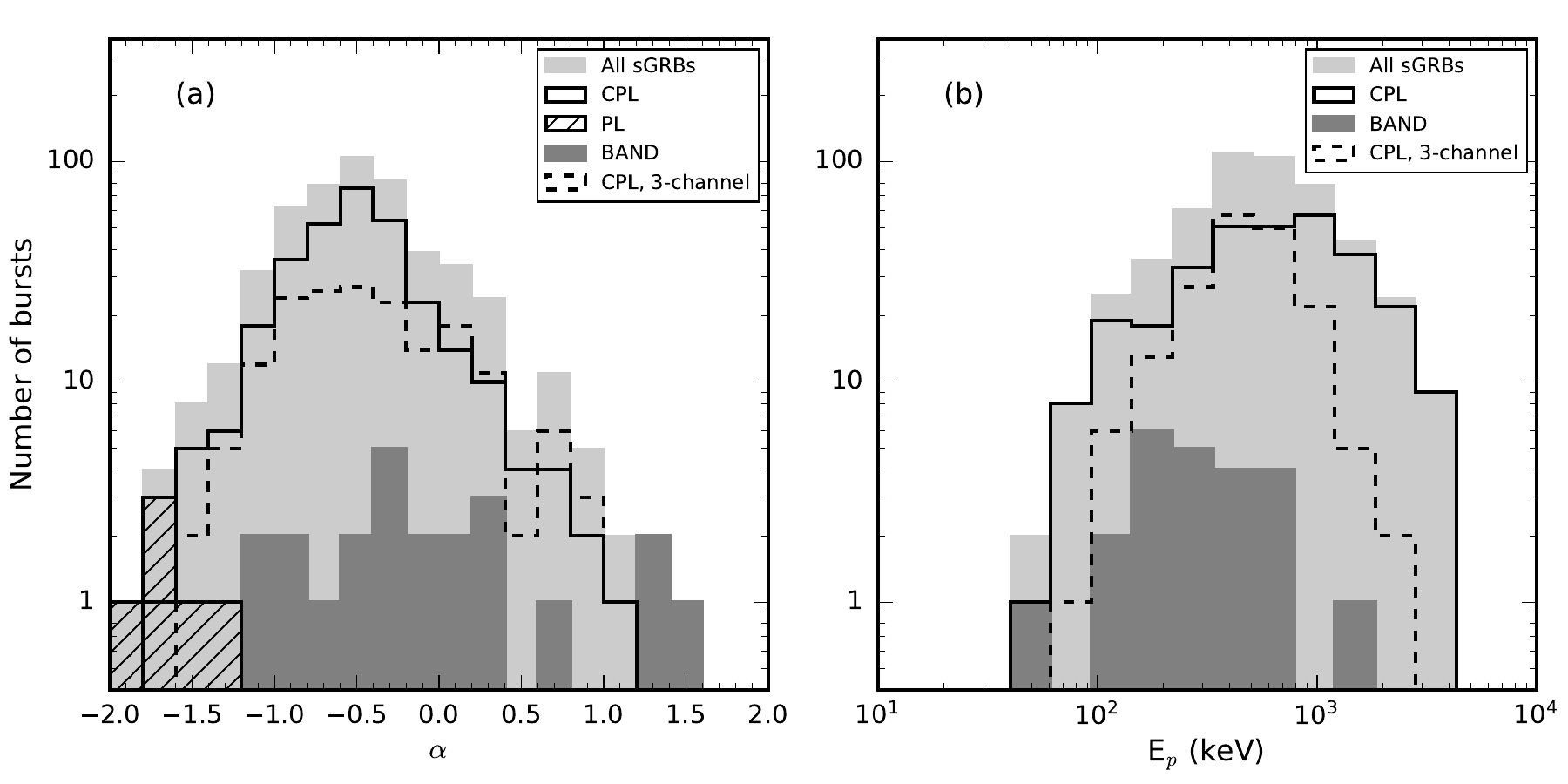}
    \caption{Distributions of $\alpha$ (left) and $E_{\rm p}$ (right) obtained for GRBs with different best-fit models. Solid black lines -- CPL fits to multi-channel spectra, dashed curves -- CPL fits to three-channel spectra, the hatched histogram displays PL indices, and dark gray histograms -- the BAND model parameters. Light gray histograms show the summed-up distributions for all sGRBs.}
    \label{fig_fit_params}
\end{figure*}

Table~\ref{tab_mult} provides the results of the multi-channel spectral analysis for the 94 TI spectra and the 11 spectra near the peak count rate of the Sample~II bursts.
The 10 columns in Table~\ref{tab_mult} contain the following information: (1) the burst designation (see Table~\ref{tab_loc}); (2) the spectrum type, where `i' indicates TI spectrum used to calculate the total energy fluence $S$; `p' indicates that the spectrum is measured near the peak count rate (and is used to calculate the peak energy flux $F_\mathrm{peak}$), or `i,p'~--- TI spectrum was used to calculate both $S$ and $F_\mathrm{peak}$; columns (3) and (4) contain the spectrum start time $T_\mathrm{start}$ (relative to $T_0$) and its accumulation
time $\Delta T$; column (5) lists models with the null hypothesis probabilities $P>10^{-6}$ for each spectrum, the best-fit model is indicated with the asterisks; columns (6)--(8) contain $\alpha$, $\beta$, and $E_\mathrm{p}$; column (9) represents the normalization (energy flux in the 10~keV--10~MeV band); column (10) contains $\chi^2/\mathrm{dof}$ along with the null hypothesis probability $P$. 
In cases where the errors for $\beta$ are not constrained, we report the upper limit on $\beta$.
The 94 best-fit models for the TI spectra include 83 CPL, 9 BAND, and 2 PL.

Table~\ref{tab_3chan} contains the results of the CPL fits for the 105 three-channel spectra of Sample~II bursts. 
The seven columns contain the following information: column (1) gives the burst designation
(see Table~\ref{tab_loc}); columns (2) and (3) contain the spectrum start
time $T_\mathrm{start}$ (relative to $T_0$) and its accumulation time $\Delta T$; column (4) provides the spectral model; columns (5) and (6) comprise $\alpha$ and $E_\mathrm{p}$, respectively; column (7) presents the normalization (energy flux in the 10\,keV--10\,MeV band).

The spectral parameter distributions for the Full sGRB Sample are presented in Figure~\ref{fig_fit_params}.
The distribution of $\alpha$ has a maximum at about -0.5 and spreads from $-2.0$ to $\sim 1.5$. 
The maximum of the $E_{\rm p}$ distribution lies between 400 and 500\,keV, and the distribution extends over two orders of magnitude from a few tens of keV up to a few MeV. 
These results are in agreement with the results obtained in S16.\footnote{To ensure the correct comparison to S16, we made CPL fits with the updated DRM to a few dozen GRB spectra from the S16 sample and obtained the spectral parameters (photon indices, peak energies, and fluxes) consistent within errors with those reported previously.} 

\subsubsection{\label{ssec_peculiarity} The peculiar GRB~111113A}
For the intense GRB20111113\_T18613 (GRB 111113A) the light curve consists of a single hard peak with the total duration of $\sim 160$\,ms.
There is a hint of a weaker emission starting $\sim 200$\,ms before the main pulse and extending to $T_0+500$\,ms.  
The burst spectrum demonstrates an excess above the hard CPL continuum ($\alpha \sim -0.7$, $E_p \sim 1500$ keV) at lower energies \citep[$\lesssim 50$\,keV;][]{Golenetskii2011}. 
The spectral excess is well described by either a black-body component with the temperature $kT \sim 7.7$\,keV or an additional PL component with the photon index $\sim 2.1$. 
The blackbody and PL components contributes $\sim$1\% and $\sim$8\% to the 10 keV--10\,MeV energy flux, respectively ($7 \times 10^{-7}$~erg~cm$^{-2}$~s$^{-1}$ for black-body and $3.6 \times 10^{-6}$~erg~cm$^{-2}$~s$^{-1}$ for PL vs. $4.6 \times 10^{-5}$~erg~cm$^{-2}$~s$^{-1}$ for CPL).

\subsection{\label{ssec_fluences_pfluxes} Fluences and Peak Fluxes}

\begin{table}
\begin{threeparttable}
\caption{Fluences and Peak Fluxes (Sample~II).}
\label{tab_fluence_pflux}
\begin{tabular}{cccc}
\toprule
\headrow Designation & $S$\tnote{a} & $T_{\rm peak}$\tnote{b} & $F_{\rm peak}$\tnote{a}  \\
 & $10^{-6}$ erg cm$^{-2}$ & (s) & $10^{-6}$ erg cm$^{-2}$ s$^{-1}$ \\
\midrule
GRB20110212\_T47551 & 0.90$_{-0.15}^{+0.23}$    & $-0.010$ & 28.8$_{-5.7}^{+7.9}$ \\
GRB20110221\_T18490 & 3.33$_{-0.62}^{+0.87}$    & $-0.006$ & 46.3$_{-12.8}^{+17.4}$ \\
GRB20110323\_T57460 & 0.57$_{-0.09}^{+0.10}$    & $-0.054$ & 14.4$_{-3.2}^{+3.4}$ \\ 
GRB20110401\_T79461 & 2.51$_{-0.70}^{+2.93}$    & $-0.070$ & 11.4$_{-4.0}^{+13.5}$ \\
GRB20110510\_T80844 & 1.20$_{-0.18}^{+0.19}$    & $-0.044$ & 4.0$_{-1.2}^{+1.2}$ \\
\bottomrule
\end{tabular}
\begin{tablenotes}[hang]
\item[]Table note
\item[a]In the 10\,keV--10\,MeV energy band.
\item[b]Relative to the trigger time. \\
(This table is available in its entirety in machine-readable form via \url{http://www.ioffe.ru/LEA/shortGRBs/Catalog3/}.)
\end{tablenotes}
\end{threeparttable}
\end{table}

\begin{figure*}
    \centering
    \includegraphics[width=0.9\textwidth]{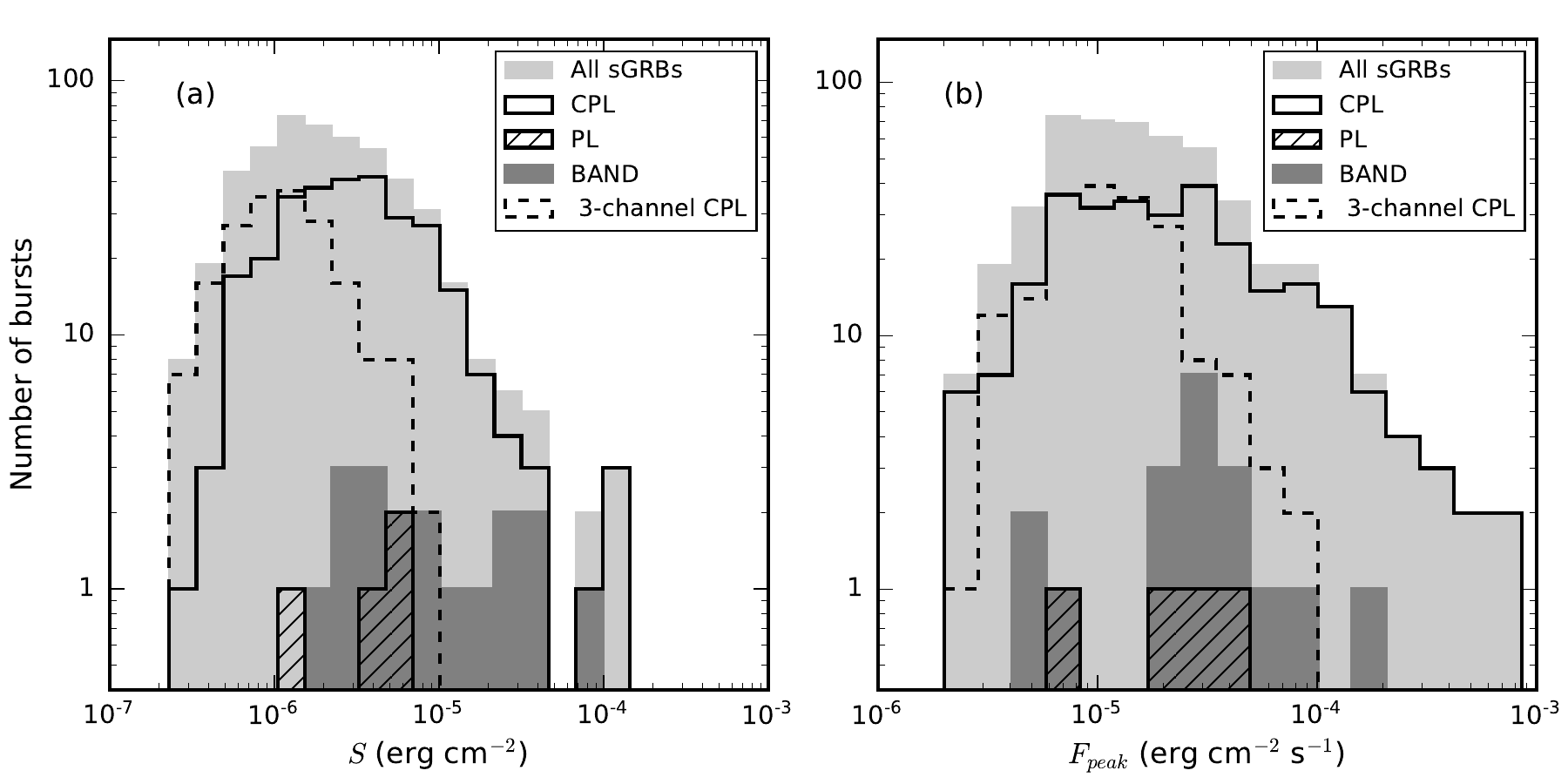}
    \caption{Flunce $S$ (left) and peak flux $F_\mathrm{peak}$ (right) distributions. Light gray histogram corresponds to all sGRBs, solid black curve represents sGRBs with multi-channel spectra fitted by a CPL model, dashed black curve represents sGRBs with three-channel spectra, hatched histogram corresponds to PL model, and dark gray histogram corresponds to BAND model.}
    \label{fig_fluence_pflux}
\end{figure*}

We derived $S$ and $F_\mathrm{peak}$ using the normalizing energy flux of the best-fit spectral model in the 10\,keV--10\,MeV band. 
In order not to overestimate the burst energetics, we used the CPL model for the cases, where a divergent PL model (i. e. with $\alpha > -2$) was chosen as the best-fit model.
Since the spectrum accumulation interval typically differs from the $T_{100}$ interval, a correction was introduced when calculating $S$, for more details see~S16 and T17.

$F_\mathrm{peak}$ was calculated as a product of the best-fit spectral model energy flux and the ratio of the 16\,ms peak count rate to the average count rate in the spectrum accumulation interval.
Typically, the peak count rate to the spectrum count rate ratio was calculated using counts in the G2+G3 light curve; the G1+G2, G2 only, and the G1+G2+G3 combinations were also considered depending on the emission hardness and signal-to-noise ratio.

For sGRBs with EE, $S$ and $F_\mathrm{peak}$ of IP and EE were estimated independently.

Table~\ref{tab_fluence_pflux} contains $S$ and $F_\mathrm{peak}$ for the 199 bursts of the Sample~II. The first column gives the burst designation (see Table~\ref{tab_loc}). 
The three subsequent columns give $S$; the start time of the 16\,ms time interval, when the peak count rate in the G2+G3 band is reached; and $F_\mathrm{peak}$. 

The distributions of $S$ and $F_\mathrm{peak}$ are shown in Figure~\ref{fig_fluence_pflux}. Both distributions extend over a few orders of magnitude,
(0.2--200)$\times 10^{-6}$~erg~cm$^{-2}$ for $S$ and (0.02--1000)$\times 10^{-5}$~erg~cm$^{-2}$~s$^{-1}$ for $F_{\rm peak}$.
The $S$ and $F_{\rm peak}$ distributions peak at about $10^{-6}$~erg~cm$^{-2}$ and $10^{-5}$~erg~cm$^{-2}$~s$^{-1}$, respectively.

\subsection{\label{ssec_ee} sGRBs with EE}

\begin{table*}
\begin{threeparttable}
\caption{sGRBs with EE (Sample~II). }
\label{tab_ee}
\begin{tabular}{ccccccccccccc}
\toprule
\headrow Designation & EE $T_{\rm 0}$\tnote{a} & EE $T_{\rm 100}$ & $T_{\rm start}$\tnote{a} & $\Delta T$ & Model\tnote{b} & $\alpha$ & $\beta$ & $E_{\rm p}$ & Fluence\tnote{c} & EE to IP\tnote{d} & $\chi^2$/dof (Prob.) \\
\midrule
GRB20111121\_T59182\tnote{e} & 1.008 &  94.288 &  &  &  &  &  &  &  &  &  \\
GRB20120624\_T26662\tnote{f} & 8.400 & 134.704 &  &  &  &  &  &  &  &  &  \\
GRB20120816\_T86298 & 1.376 & 18.512 & 0.768 & 16.384 & *PL & $-1.63_{-0.14}^{+0.14}$ &  &  &  &  & 50.3/61 (8.3e-1) \\
 &  &  &  &  & CPL & $-0.47_{-0.85}^{+1.20}$  &  & $224_{-58}^{+258}$ & $2.08_{-0.51}^{+1.15}$ & 0.02 & 47.6/60 (8.8e-01) \\
GRB20170127\_T05744 & 3.696 & 88.912 & 0.256 & 90.112 & *PL & $-1.84_{-0.09}^{+0.08}$ &  &  &  &  & 53.2/59 (6.9e-1) \\
& & & & & CPL & $-1.65_{-0.20}^{+0.24}$ & & $257_{-99}^{+2722}$ & $10.22_{-1.83}^{+7.05}$ & 1.03 & 52.1/58 (6.9e-01) \\
& & & & & BAND & $-1.64_{-0.20}^{+0.24}$ & <-1.86\tnote{g} & $250_{-95}^{+409}$ & & & 52.1/57 (6.6e-01) \\
GRB20170304\_T20449 & 2.144 & 56.288 & 8.704 & 49.408 & CPL & $-1.14_{-0.02}^{+0.02}$ & & $1118_{-64}^{+69}$ & & & 137.4/98 (5.4e-03) \\
& & & & & *BAND & $-1.11_{-0.02}^{+0.02}$ & $-2.38_{-0.26}^{+0.16}$ & 975$_{-82}^{+89}$ & 202.77$_{-9.12}^{+8.87}$ & 1.85 & 130.1/97 (1.4e-2) \\
GRB20180618\_T02591 & 0.608 & 3.184 & 0.256 & 8.192 & *PL & $-1.84_{-0.29}^{+0.21}$  &  &  &  &  & 95.9/78 (8.2e-2) \\
& & & & & CPL & $-0.24_{-0.80}^{+2.27}$ & & $187_{-103}^{+97}$ & $0.79_{-0.45}^{+0.50}$ & 0.11 & 90.5/77 (1.4e-01) \\
& & & & & BAND & $-0.22_{-0.82}^{+1.36}$ & <-2.49\tnote{g} & $187_{-104}^{+96}$ & & & 90.5/76 (1.2e-01) \\
\bottomrule
\end{tabular}
\begin{tablenotes}[hang]
\item[]Table note
\item[a]Relative to the trigger time.
\item[b]The best-fit model is indicated by the asterisk.
\item[c]In units of $10^{-6}$ erg cm$^{-2}$.
\item[d]EE to IP fluence ratio.
\item[e]EE was observed only in G1 channel, spectral fit is not feasible.
\item[f]EE was observed only in G2 channel, spectral fit is not feasible.
\item[g]Upper limits.
\end{tablenotes}
\end{threeparttable}
\end{table*}

The parameters of the extended emission of extended emissions of six Sample II sGRBs are listed in Table~\ref{tab_ee}. 
For four of these bursts the EE was intense enough to perform spectral fitting (see, Section~\ref{ssec_spec_analysis}). 
The 12 columns of Table~\ref{tab_ee} contain the following information:
(1) the burst designation; (2) the EE start time relative to $T_0$; (3) the EE total duration $T_{100, \mathrm{EE}}$; (4) the EE TI spectrum start time $T_{\rm start, EE}$ and (5) its accumulation time $\Delta T$; (6) models with the null hypothesis
probabilities $P > 10^\mathrm{-6}$ (the best-fit model is marked by an asteriks); (7)-(9) the model parameters: $\alpha$, $\beta$, $E_{\rm p}$; (10) the energy fluence in the 10\,keV--10\,MeV range; (11) EE to IP fluence ratio; and (12) the $\chi^2$/dof along with the null hypothesis probability. 
As for IPs, for the cases where a divergent PL model was chosen as the best-fit model, we used a CPL model for fluence calculation.

Three out of four bursts with constrained spectral fits are best described by PL and one~--- by BAND. 
For these models, the best-fit photon indices $\alpha$ are in the range between $-1.84$ and $-1.11$. 
The peak energy of the EE fitted by BAND is rather hard, 975\,keV, but this value is softer than that for the IP. 
For two bursts, the EE fluence is significantly smaller than that of the IP, and for the other two bursts, the EE fluence is comparable to the IP fluence. 

In S16, 30 sGRBs with EE, or $\sim 10$\% of Sample~I, were reported, among which 21 bursts were bright enough to produce reasonable spectral fits. 
In this work, for the more recent Sample~II, we found only six (3\%) bursts with EE. 
The lower fraction of bursts with EE in Sample~II may be due to the shift of the KW energy range to higher energies, which makes rather faint and soft EE harder to detect.

\section{\label{sec_discussion}Discussion}

\subsection{\label{ssec_disc_dist} Burst parameters and classification}

\begin{table*}
\begin{threeparttable}
\caption{Parameter distributions for different sGRB Types  in the Full sGRB Sample\tnote{a}.}
\label{tab_dist_ci}
\begin{tabular}{lccccc}
\toprule
\headrow Parameter & Type I & Type II & Type Iee & All sGRBs\tnote{b} & MGFs \\
\midrule
Number of bursts & 357 & 44 & 26 & 487 & 5 \\
\hline
$\tau_\mathrm{lag21}$ (ms) & 2 [-16, 45] & 34 [-13, 139] & 2 [-7, 34] & 4 [-15, 88] & 0 [-2, 3] \\
$\tau_\mathrm{lag32}$ (ms) & 2 [-22, 34] & 42 [8, 113] & 6 [-33, 48] & 3 [-23, 43] & 0 [-3, 1] \\
$\tau_\mathrm{lag31}$ (ms) & 2 [-24, 67] & 86 [18, 253] & 8 [-25, 73] & 5 [-24, 91] & 4 [-3, 27] \\
\hline
$\delta T$ (ms) & 74 [8, 386] & 267 [59, 1246] & 18 [2, 138] & 82 [8, 615] & 2 [2, 28] \\
$\tau_\mathrm{rise}$ (ms) & 100 [11, 569] & 488 [90, 1189] & 224 [8, 651] & 129 [13, 688] & 5 [3, 23] \\
$\tau_\mathrm{decay}$ (ms) & 140 [23, 937] & 692 [244, 2754] & 423 [23, 12069] & 205 [24, 1420] & 135 [41, 304] \\
$\tau_\mathrm{rise}$/$\tau_\mathrm{decay}$ & 0.52 [0.05, 10.07] & 0.50 [0.12, 2.60] & 0.51 [0.01, 9.17] & 0.47 [0.05, 9.33] & 0.10 [0.01, 0.27] \\
\hline
$\alpha$ & -0.43 [-1.02, 0.54] & -0.72 [-1.58, 0.03] & -0.60 [-1.03, 0.21] & -0.50 [-1.20, 0.53] & -0.12 [-0.89, 0.64] \\
$\beta$ & -2.0 [-2.7, -1.7] & -2.6 [-2.8, -2.0] &  & -2.2 [-3.0, -1.8] &  \\
$E_p$ (keV) & 619 [254, 1784] & 121 [68, 200] & 1383 [302, 2724] & 519 [121, 1889] & 817 [321, 1649] \\
Fluence\tnote{c} & 1.9 [0.5, 11.1] & 1.7 [0.4, 8.8] & 6.9 [1.0, 107.8] & 2.0 [0.5, 14.6] & 7.3 [1.1, 25.7] \\
Peak flux\tnote{d} & 16 [6, 93] & 5 [3, 22] & 62 [8, 435] & 14 [4, 110] & 275 [43, 740] \\
\bottomrule
\end{tabular}
\begin{tablenotes}[hang]
\item[]Table note
\item[a]For each parameter the median value is given followed by 90\% CI in square brackets.
\item[b]Except MGFs.
\item[c]In units of $10^{-6}$~erg~cm$^{-2}$.
\item[d]In units of $10^{-6}$~erg~cm$^{-2}$~s$^{-1}$.
\end{tablenotes}
\end{threeparttable}
\end{table*}

We have presented the results of the systematic spectral and temporal analysis of 199 short GRBs, detected by KW during 2011--2021, which extends the KW sGRB sample to 494 events (1994--2021; in 27 years of operation). This implies the sGRB detection rate by KW of $\sim 19$ per year.
Among GRB experiments \citep[see][for a recent review]{Tsvetkova2022}, the KW sGRB sample is one of the largest to date, \textit{Swift}-BAT has detected 132 sGRBs up to 2021, August \footnote{\url{https://swift.gsfc.nasa.gov/results/batgrbcat/}}~\citep[$\sim 8$ per year;][]{Lien_2016ApJ_829_7}, \textit{AGILE}-MCAL has detected $\sim 220$ sGRBs from 2007 to 2020 \citep[$\sim 17$ per year;][]{Ursi2022}, and \textit{Fermi}-GBM has detected $\sim 600$ up to 2021, August\footnote{\url{https://heasarc.gsfc.nasa.gov/W3Browse/fermi/fermigbrst.html}}~\citep[$\sim 37$ per year;][]{von_Kienlin_2020ApJ_893_46}. 
In comparison with the Fermi-GBM sample the KW bursts represent the brighter end of the sGRB population, with energy fluences above $0.3\times 10^{-6}$~erg~cm$^{-2}$ in the 10\,keV--10\,MeV energy range. 

\begin{figure*}
    \centering
    \includegraphics[width=0.9\textwidth]{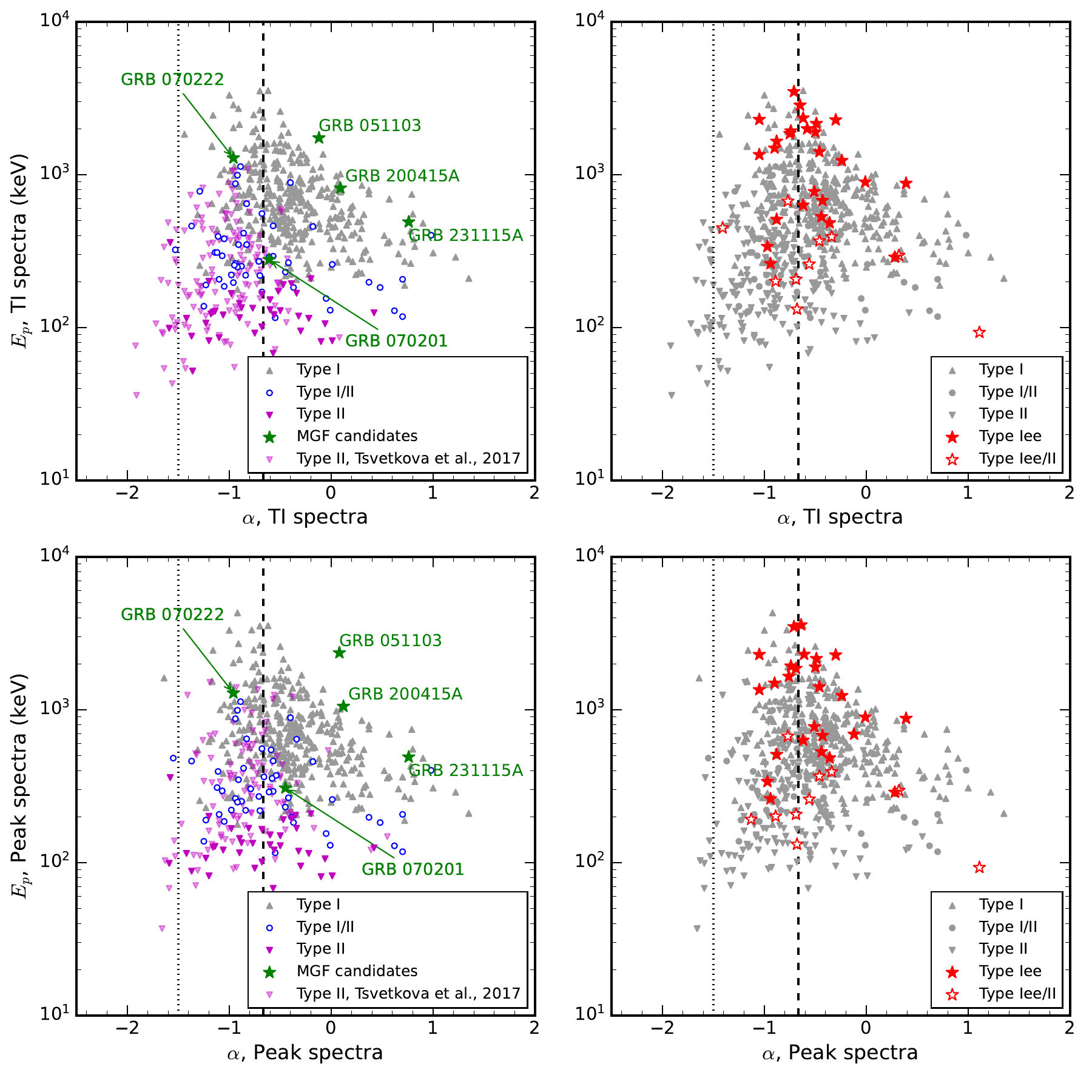}
    \caption{Distributions of $\alpha$ and $E_{\rm p}$ for TI spectra (upper panels) and peak spectra (lower panels) for different types. Left panels: Type~I bursts (gray triangles), Type~I/II bursts (empty blue circles), Type~II bursts (magenta triangles), Type~II bursts from T17 (light magenta triangles), and MGF candidates (green stars). Right panels: Type~Iee bursts (red stars), Type~Iee/II bursts (empty red stars), other types are shown in gray. Dotted and dashed vertical lines refer to the synchrotron fast-cooling and the synchrotron slow-cooling limit respectively.}
    \label{fig_alpha_vs_ep}
\end{figure*}

Table~\ref{tab_dist_ci} contains median values and 90\% CIs of sGRB parameter distributions calculated for the Full sGRB Sample for different burst types: Type~I, Type~II, Type~Iee, MGFs, and all sGRBs, excluding MGFs. 
Most Type~I, Type~Iee bursts and MGFs have negligible spectral lags within the uncertainties. 
MGFs differ from other sGRBs by considerably shorter $\delta T$ and $\tau_{\rm rise}$ scales, and a prominent asymmetry of the pulse profile that is characterized by a low $\tau_{\rm rise}$-to-$\tau_{\rm decay}$ ratio.
Type II bursts of our sample represent a tail of the whole Type II GRB population. 
They,  unlike other types of sGRBs, display some typical properties of long GRBs, e. g., significant positive spectral lags and longer $\delta T$ scales.

In the lower section of Table~\ref{tab_dist_ci}, we characterize the spectral parameter distributions for the TI spectra. 
The median value of the photon index $\alpha$ for all sGRBs is close to $-1/2$, and,  within errors, only four sGRBs have $\alpha$ values steeper than $-3/2$,
violating the synchrotron fast-cooling limit \citep{Preece1998}. 
It should be noted that harder values of $\alpha$ tend to have larger fit uncertainties.
For 160 sGRBs (32\% out of Full sGRB Sample), 90\% CI of $\alpha$ are above $-2/3$, thus violating the synchrotron slow-cooling limit \citep{Sari1998}. 
Ten sGRBs, within uncertanties, are characterized by positive values of $\alpha$, which can be related to the admixture of thermal emission to the synchrotron continuum \citep{Axelsson2012, Burgess2015}.
The IPs of sGRBs with EE are in general harder spectrally and more intense as compared to the ``genuine'' Type~I bursts. 
This may partly be due to a selection effect that results in the possible EEs of weaker Type I bursts being below the KW detection threshold.

MGFs are characterized by higher $E_\mathrm{p}$ values and much higher $F_\mathrm{peak}$ than Type~I bursts.
For two MGFs (GRB\,051103 and GRB\,200415A), $\alpha$ of the TI spectra is close to zero, and one MGF (GRB\,231115A) is characterized by a positive $\alpha$ value which can be explained by a growing contribution of a thermal spectral component during the decaying phase of the burst \citep{Svinkin2021}.
For two other MGFs (GRB\,070201 and GRB\,070222), $\alpha$ lies, within uncertanties, between fast- and slow- synchrotron cooling limits.

The majority ($\sim 92$\%) of sGRBs with multi-channel spectra are best-fitted by CPL, while BAND is the best-fit model only for 18 out of 308 ($\sim 6$\%) bursts with multi-channel spectra. 
Among the latter bursts there are nine Type~I bursts, four Type~II bursts, three Type~I/II bursts, one Type~Iee burst, and the MGF candidate GRB\,051103. 
The low fraction of BAND spectra, as compared to long GRB population (e.g., $\sim$38\% in the T17 sample), could be attributed to relatively poor count statistics in short bursts that do not allow to constrain properly their spectral shapes at higher energies. 
On the other hand, the cutoff spectra could be a distinct feature of merger-origin GRBs. 
Recently, a number of long GRBs have been confirmed to originate from compact object mergers that underlines the ambiguity of GRB classifications based on duration only. For at least one of them, the ultra-bright GRB\,230317A, accompanied by a kilonova \citep{Levan2024}, high signal-to-noise spectra are best described by CPL during the whole main emission phase (Svinkin, D.~S., et al., in preparation). 
Thus, the statistically-significant exponential cut-off at higher energies may point to the merger origin for some long-duration GRBs.

\begin{figure*}
    \centering
    \includegraphics[width=0.9\textwidth]{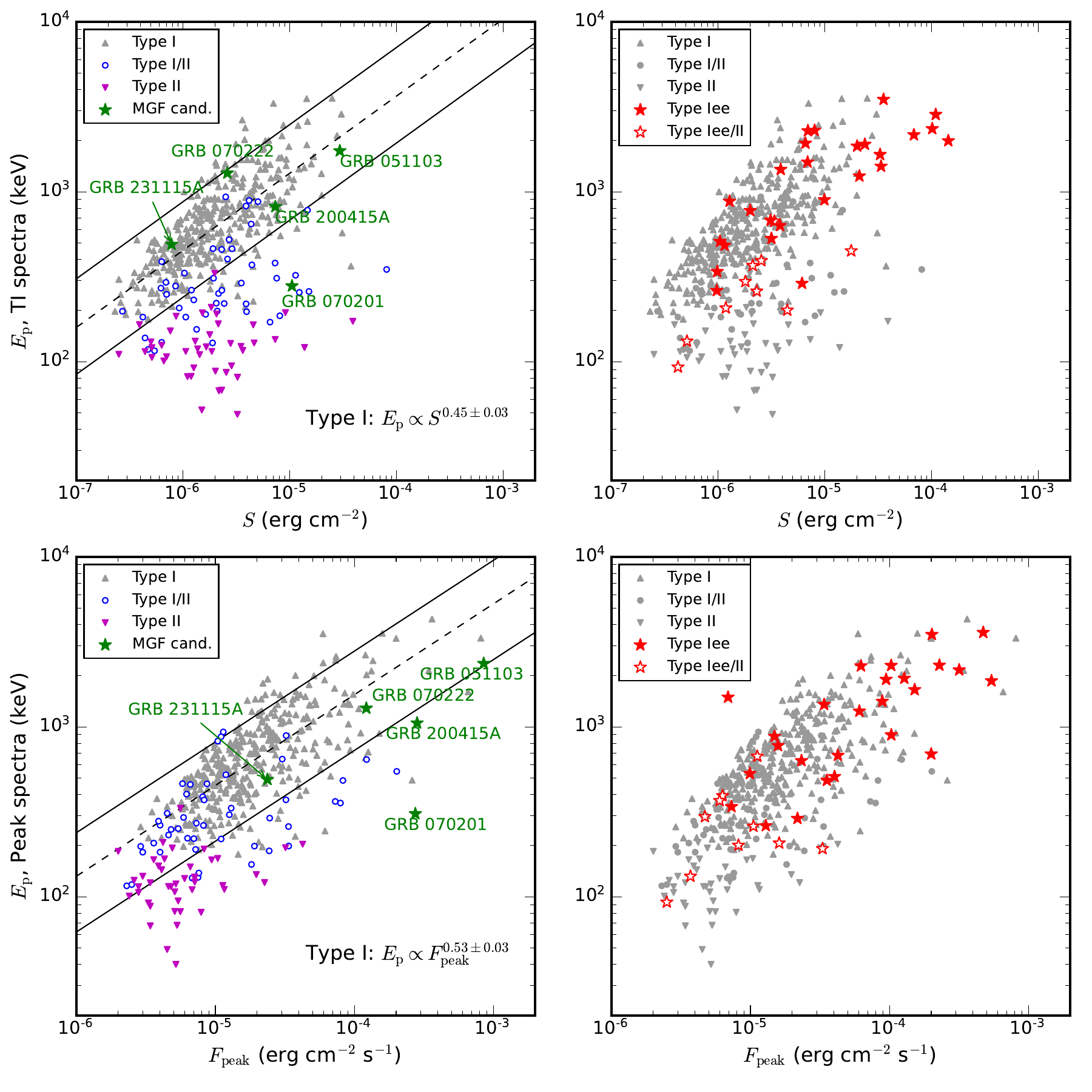}
    \caption{$E_{\rm p}$--$S$ (upper panels) and $E_{\rm p}$--$F_{\rm peak}$ (lower panels) distributions for different sGRBs types. Left panels: Type~I (gray triangles), Type~I/II (empty blue circles), Type~II (magenta triangles), MGF candidates (green stars); regressions for Type~I bursts (black dashed line) and 90\% PIs for Type~I bursts (black solid lines). Right panels: Type~Iee (red stars), Type~Iee/II (empty red stars), other types are shown in gray. 
     }
    \label{fig_ep_vs_en}
\end{figure*}

Figure~\ref{fig_alpha_vs_ep} presents 2D distributions of $\alpha$ and $E_{\rm p}$ for different GRB types, including Type~II bursts from T17. 
The general form of the distributions is similar to that reported by \cite{Poolakkil2021} for GRBs detected by \textit{Fermi}-GBM. 
It should be noted that the spectra with lower $E_{\rm p}$ values tend to have larger uncertainties in $\alpha$, which may explain the increase in the width of the index distribution with decreasing $E_{\rm p}$.
The $\alpha$--$E_{\rm p}$ distributions for IPs of sGRBs with EE are generally consistent with those for Type~I burst. 
A vivid difference is observed between Type~I and Type~II bursts: Type~II bursts have in general lower $E_{\rm p}$ values than Type~I bursts and are characterized by significantly softer $\alpha$ indices for both TI and peak spectra. 
As with the hardness-duration relation, there is no distinct boundary between GRBs of different Types in the $\alpha$--$E_{\rm p}$ plane, but the typical parameter values for Type~I GRBs in this plane tend to be different from those for Type~II bursts. 
To compare the burst distributions in the $\alpha$--$E_{\rm p}$ plane we applied the bivariate Kolmogorov-Smirnov test \citep[2D KS-test,][]{Peacock1983, Fasano1987, Press1992}.
We found that the probability that $\alpha$--$E_{\rm p}$ distributions are similar for Type~I and Type~II bursts is negligible ($P<10^{-20}$), while the distributions for Type~I bursts and Type~Iee bursts are rather consistent: 2D KS-test results in probability $P= 6 \times 10^{-3}$ for TI spectra and $P= 3 \times 10^{-2}$ for peak spectra.

Figure~\ref{fig_ep_vs_en} shows $E_{\rm p} - S$ and $E_{\rm p} - F_\mathrm{peak}$ relations for the Full sGRB Sample. 
To estimate slopes $m$ (i.e. $E_{\rm p} \propto S^m$) for these relations, we used the fitting method with bivariate correlated errors and intrinsic scatter \citep[BCES,][]{Akritas1996} realized in Python \citep{Nemmen2012}. 
For Type~I bursts we found $E_{\rm p, I} \propto S_{\rm I}^{\rm 0.45 \pm 0.03}$ and $E_{\rm p, I} \propto F_{\rm peak, I}^{0.53 \pm 0.03}$. 
Type II sGRBs, which represent a low-intensity fraction of long-soft GRB population, lie below the lower bound of 90\% prediction interval (PI) of $E_{\rm p} - S$ relation for Type I sGRBs. 
This result is consistent with studies of GRB rest-frame properties \citep[T17;][]{Zhang2018, Minaev2020, Tsvetkova2021}, which demonstrate that populations of Type I and Type II GRBs occupy different regions in the intrinsic $E_\mathrm{p}$ -- isotropic burst energy release ($E_\mathrm{iso}$) plane.

Four out of five MGF candidates fall into or are very close to the 90\% PI on both  $E_{\rm p}$--$S$ and $E_{\rm p}$--$F_{\rm peak}$ planes. 
The remaining MGF candidate, GRB\,070201, is an outlier with much lower peak energy as expected for Type~I bursts. 
Two right panels of Figure~\ref{fig_ep_vs_en} show the 2D parameter distributions for IPs of sGRBs with EE. For Type Iee bursts the distributions do not differ significantly from those for genuine Type I GRBs.

\subsection{\label{ssec_disc_en_cum}sGRB energetics and spatial distribution}
\begin{figure*}
    \centering
    \includegraphics[width=0.95\textwidth]{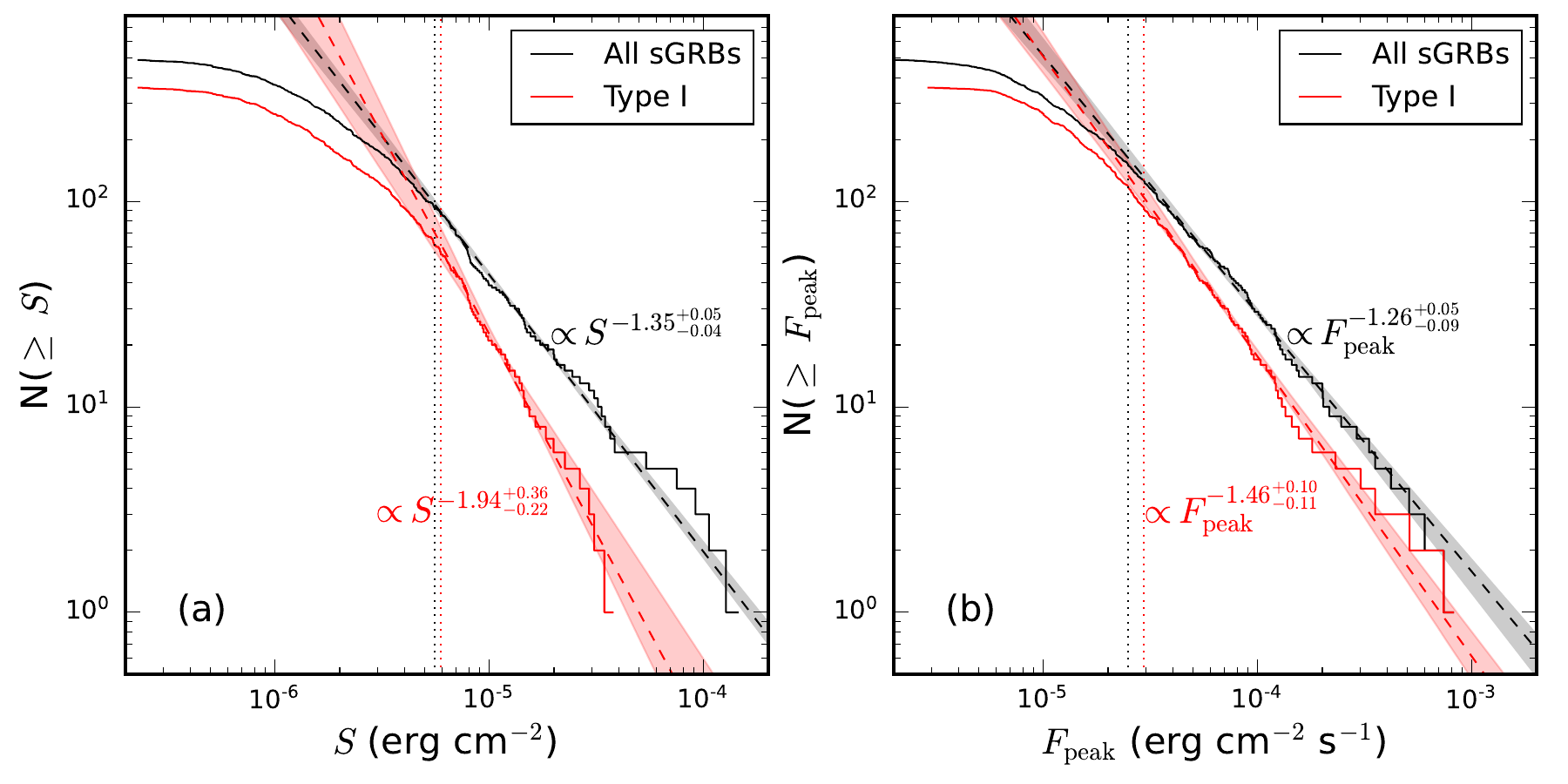}
    \caption{Cumulative distributions of the total energy fluence (left panel) and the peak energy flux (right panel). In both panels, black and red histograms represent cumulative distributions for the whole sGRB sample and for the Type~I GRB sub-sample, respectively. Dashed lines indicate PL fit to the distributions for all sGRBs (black) and Type~I bursts (red). Dotted vertical lines mark left boundaries of the ranges, where the assumption that the data are PL-distributed is valid. Gray and pink areas on both plots indicate 90\% CI for PL fit for all sGRBs and Type~I bursts, respectively.}
    \label{fig_en_cum}
\end{figure*}
 

Figure~\ref{fig_en_cum} presents the cumulative distributions of fluences and peak fluxes for Full sGRB Sample (excluding MGF candidates) and Type~I subsample. 
Due to various instrumental biases, which result in the lack of the faint bursts in the sample, both log$N$--log$S$ and log$N$--log$F_{\rm peak}$ distributions follow a PL in a limited range of fluences and peak fluxes.
We used \texttt{powerlaw} Python package~\citep{Alstott2014} both for the determination of the range, where a PL is valid, and for the estimation of the PL slope $\gamma$. 
We performed Monte-Carlo simulations of multiple $S$ and $F_{\rm peak}$ sets, for each set we estimated the range for PL and PL slope. 
The resulting slope was taken as the median value for all simulated sets, and the 68\% CI for the slope was estimated.

The log$N$--log$S$ distribution for all sGRBs follows a PL for fluences $\gtrsim 5 \times 10^{-6}$~erg~cm$^{-2}$, with the slope $\gamma_{\rm all, S}=-1.35_{\rm -0.04}^{\rm +0.05}$.
The obtained slope is flatter than -3/2 (the slope for homogeneously distributed sources), which is expected for red-shifted sources \citep{Meegan1992}. 
The distribution for Type~I bursts only obey a PL for fluences above $\sim 6 \times 10^{-6}$~erg~cm$^{-2}$, with the slope $\gamma_{\rm I, S}=-1.94_{\rm -0.22}^{\rm +0.36}$, which is steeper, than expected for homogeneously distributed sources. 
The lack of high-fluence Type~I sources can be related to the absence of Type~I bursts with longer durations in our sample. 
Some of such longer Type~I GRBs lie within the short-hard GRB cluster, but do not obey our criterion of short bursts, and some merger-origin GRBs could, in fact, be long GRBs \citep[see, e.g.,][]{Petrosian2024}.
For the log$N$--log$F_{\rm peak}$ distribution, a PL is applicable for the peak fluxes $ \gtrsim 2 \times  10^{-5}$~erg~cm$^{-2}$~s$^{-1}$ for all sGRBs and $ \gtrsim 3 \times  10^{-5}$~erg~cm$^{-2}$~s$^{-1}$ for Type~I bursts.
The PL slopes constitute $\gamma_{\rm all, F}=-1.26_{\rm -0.09}^{\rm +0.05}$ for all sGRBs and $\gamma_{\rm I, F}=-1.46_{\rm -0.11}^{\rm +0.10}$ for Type~I bursts.
The value of $\gamma_{\rm I, F}$ is close to -3/2 which is related to low red shifts of Type~I  bursts. 

\section{\label{sec_conclusions}Summary and conclusions}

In this catalog, we analyze the temporal and spectral characteristics as well as energetics of 199 sGRBs, detected by KW between January 1, 2011 and August 31, 2021. 
Together with the Second Catalog, the total KW sGRB sample comprises 494 sGRBs detected in 1994-2021 and the recent MGF candidate GRB\,231115A.
For the whole set of 494 sGRBs, we presented statistical distributions of spectral lags, minimum variability time scales, rise and decay times, spectral fitting parameters, energy fluences, and peak energy fluxes.
We compared distributions of these characteristics for different sGRBs subsamples: Type~I (merger-origin), Type~II (collapsar-origin), Type~Iee (merger-origin bursts with extended emission), and MGF candidates.
We found that MGF candidates differ from other sGRBs by shorter variability time scales and rise times, and the higher asymmetry of the emission pulse. 
Type Iee burst IPs demonstrate properties similar to genuine Type I bursts, suggesting these bursts represent the more intense and spectrally hard fraction of the Type~I population.
Compared to other sGRBs from our sample, Type~II bursts are characterized by significant positive spectral lags, longer variability timescales, and softer energy spectra, all of which are typical of long GRBs of collapsar origin.

\begin{acknowledgement}
The authors thank the software developers: 
matplotlib \citep{Hunter2007}; 
XSPEC \citep{Arnaud1996};
Astropy \citep{Astropy_2013AandA, Astropy_2018AJ};
BCES \citep{Akritas1996, Nemmen2012};
healpy (\citet{Gorsky_2005ApJ_622_759, Zonca_2019JOSS_4_1298}, \url{https://healpix.sourceforge.io}); 
mhealpy (\cite{Martinez2022} \url{https://mhealpy.readthedocs.io}); 
Powerlaw \citep{Alstott2014}.
\end{acknowledgement}

\paragraph{Funding Statement}
A.E.T. acknowledges support from the HERMES Pathfinder–Operazioni 2022-25-HH.0.

\paragraph{Competing Interests}
None

\paragraph{Data Availability Statement}
The data underlying this article will be shared on reasonable request to the corresponding author.
Tables~\ref{tab_loc}--\ref{tab_fluence_pflux} in machine-readable form and plots of the KW sGRBs time histories and spectral fits can be found at the Ioffe Web site \url{http://www.ioffe.ru/LEA/shortGRBs/Catalog3/}.
The current GRB list is available at \url{http://www.ioffe.ru/LEA/kw/triggers/}.

\printendnotes

\bibliography{ShortGRBs}



\end{document}